\documentclass[modern]{aastex61}
\usepackage{graphicx}
\usepackage{amsmath}
\usepackage{amssymb}
\usepackage{enumitem}

\newcommand\mybar{\kern1pt\rule[-\dp\strutbox]{.8pt}{\baselineskip}\kern1pt}

\setlist[itemize]{noitemsep, topsep=0pt, leftmargin=*}

\shorttitle{PBHs in the Solar System}
\shortauthors{Loeb}

\begin{document}

\title{Excluding Primordial Black Holes as Dark Matter Based on Solar
  System Ephemeris}

\author{Abraham Loeb}
\affiliation{Astronomy Department, Harvard University, 60 Garden
  St., Cambridge, MA 02138, USA}

\begin{abstract}
Current cosmological constraints allow primordial black holes (PBHs)
to constitute dark matter in the mass range of $10^{18}$--$10^{22}$~g.
I show that a major portion of this logarithmic window can be
ruled-out based on the Solar System ephemeris, given that the external
mass enclosed within $50$~au from the Sun did not change by more than
$\sim 5\times 10^{-14}~M_\odot~{\rm yr^{-1}}$ in recent decades.
\end{abstract}

\bigskip\bigskip
\section{Introduction}

Current cosmological constraints allow for the possibility that dark
matter is made of primordial black holes (PBHs) in the mass range of
$\sim 10^{18}$-$10^{22}~{\rm
  g}$~\citep{1974MNRAS.168..399C,2021arXiv211002821C,2024NuPhB100316494G,2024arXiv240605736C}.

Recently, \citet{2021A&A...647A.141P} used data in the Solar System
ephemeris EPM2019 to constrain the change in the mass of the Sun based
on the dynamics of Solar System objects out to $\sim 500~{\rm
  au}$. EPM2019 incorporates full 3D position and velocity vectors of
the Sun, the Moon, the eight major planets, Pluto, the three largest
asteroids (Ceres, Pallas, and Vesta) and four transneptunian objects
(Eris, Haumea, Makemake, and Sedna), covering data over more than
400~yr.

Accounting for the known components of mass loss from the Sun in
radiation or solar wind and the small mass gain from infall,
\citet{2021A&A...647A.141P} derived the following $3\sigma$ limits on
the rate of unaccounted-for mass change,
\begin{equation}
-2.9\times 10^{-14}< {\dot{\delta M}\over M_\odot}<+4.6\times 10^{-14}~{\rm per~yr}~,
\label{eq:1}
\end{equation}
where $\delta M=(M-M_\odot)$ corresponds to any mass deficit or excess
relative to the known mass budget of the Sun.

If dark matter is made of PBHs, then the temporary passage of a PBH
through the inner Solar System would introduce a transient $\delta M$
in the gravitational mass affecting all objects orbiting the Sun
outside of the PBH-Sun separation.  Here, I study the constraints set
by equation~(\ref{eq:1}) on the abundance of PBHs in the mass range of
$10^{18}$-$10^{22}$~g. In our analysis, I ignore the possibility of a
time dependent Newton's constant, because it is unlikely that such
variations would compensate random $\delta M$ fluctuations introduced
by PBHs as they enter and exit a perihelion distance of $\sim 50~{\rm
  au}$ over timescales of years. Other recent papers addressed
complementary ways for constraining PBHs from dynamical data in the
Solar
System~\citep{2023arXiv231217217T,2023arXiv231214520B,2024arXiv240314397C}.

\section{New Solar System Constraints}

Based on the latest Galactic data, the dark-matter near the Sun has a
mass-density~\citep{2022MNRAS.511.1977S,2024JCAP...08..022S},
\begin{equation}
\rho_{\rm dm}= 7(\pm 1)\times 10^{-25}~{\rm g~cm^{-3}}~,
\label{eq:2}
\end{equation}
a 3D velocity dispersion of $280(\pm 19)~{\rm km~s^{-1}}$, and a most
  probable speed relative to the Sun of,
\begin{equation}
  v=  257(\pm 11)~{\rm km~s^{-1}}~,
\label{eq:3}
\end{equation}
with a sharp truncation above $470~{\rm km~s^{-1}}$. If PBHs of a
given mass, $m=m_{20}\times 10^{20}~{\rm g}$, make the dark matter,
then their local number density is derived from equation~(\ref{eq:2}),
\begin{equation}
n=\left({\rho_{\rm dm}\over m}\right) \approx 2.4\times 10^{-5}~{\rm
  au^{-3}}m_{20}^{-1}~.
\label{eq:4}
\end{equation}

The rate by which PBHs of mass $m$ enter a volume of radius $r$ around
the Sun is given by,
\begin{equation}
  \Gamma= n\times \left(\pi r^2\right) \times v~.
  \label{eq:5}
\end{equation}
Substituting $v$ from equation~(\ref{eq:3}) and $n$ from
equation~({\ref{eq:4}) yields an entry rate,
\begin{equation}
  \Gamma= 10.2~m_{20}^{-1} \left({r \over 50~{\rm au}}\right)^2~{\rm
    yr^{-1}}~.
  \label{eq:5}
\end{equation}
For our fiducial detection volume, I consider a sphere defined by
transneptunian objects around $r\sim 50~{\rm au}$ in the EPM2019 data
which was used to derive equation~(\ref{eq:1}). For generality, I
also express our PBH constraints as a function of the bounding value
of $r$.

Multiplying the PBH entry rate in equation~(\ref{eq:5}) by the PBH
mass $m$ yields the rate by which the mass interior to a radius $r$
changes as a result of the crossing of a single PBH within that
radius from the Sun,
\begin{equation}
\dot{m}\equiv m\Gamma= 5\times 10^{-13} \left({r \over 50~{\rm
    au}}\right)^2~M_\odot~{\rm yr^{-1}}~,
  \label{eq:6}
\end{equation}
implying that for $r\sim 50~{\rm au}$ a single PBH with $m_{20}>0.1$
can violate the limits in equation~(\ref{eq:1}).

The crossing time of a radius $r$ by a PBH is given by,
\begin{equation}
\delta t=\left({r\over v}\right)= 0.93~{\rm yr}~\left({r \over 50~{\rm
    au}}\right)~,
\label{eq:7}
\end{equation}
introducing a fluctuation $\delta M$ on a relevant timescale to be
detectable in the EMP2019 data.

At any given time, the number of PBHs within the sphere of radius $r$
is,
\begin{equation}
  N=n\times \left({4\pi r^3\over 3}\right)= 12.6~m_{20}^{-1}~\left({r
      \over 50~{\rm au}}\right)^3~.
    \label{eq:8}
\end{equation}
Poisson fluctuations over a time $\delta t$ in the enclosed mass of
PBHs yield,
\begin{equation}
  \dot{\delta M}={\sqrt{N} m\over \delta t}=
   1.9\times 10^{-13} m_{20}^{1/2}~\left({r \over 50~{\rm
    au}}\right)^{1/2}~M_\odot~{\rm yr^{-1}}~,
  \label{eq:9}
\end{equation}
with a weak square-root dependence on $m$ and
$r$. Equation~(\ref{eq:9}) holds for $N>1$, namely
$r>22m_{20}^{1/3}~{\rm au}$.

Equations~(\ref{eq:6}-\ref{eq:9}) imply that the $3\sigma$ limits in
equation~(\ref{eq:1}) exclude PBHs as dark matter in the previously
allowed mass range of $6\times 10^{18}~{\rm g}<m<10^{22}~{\rm g}$ for
$r\sim 50~{\rm au}$ and the entire range of $10^{18}$-$10^{22}~{\rm
  g}$ for Sedna's semimajor axis at $r\sim 500~{\rm au}$ . At the
upper end of this mass range, a PBH with $m\sim 10^{22}~{\rm g}$ is
expected to get within $50~{\rm au}$ from the Sun once per decade and
within $\sim 8~{\rm au}$ once per 400 years. At the lower mass end,
there are $\sim 210$ PBHs with $m\sim 6\times 10^{18}~{\rm g}$ within
50~au from the Sun at any given time. The nearest is $\sim 8.4~{\rm
  au}$ from the Sun at any given time, but during 400 years the
nearest arrives as close as $\sim 0.2~{\rm au}$ at perihelion.

\section{Discussion}

I have found that the dynamical constraints from the Solar System
ephemeris EPM2019 exclude a substantial portion of the allowed
logarithmic window for PBHs as dark matter, $10^{18}$-$10^{22}$~g,
depending on the choice of the boundary radius $r$ out to which the
interior mass is not allowed to change by more than $5\times
10^{-14}~M_\odot~{\rm yr^{-1}}$. Detailed simulations of how PBHs with
a broad mass distribution across this range affect the specific
details of the EMP2019 data, are required to refine these constraints.

\bigskip
\bigskip
\bigskip
\bigskip
\section*{Acknowledgements}

This work was supported in part by Harvard's {\it Black Hole
  Initiative}, which is funded by grants from JFT and GBMF.

\bigskip
\bigskip
\bigskip

\bibliographystyle{aasjournal}
\bibliography{t}
\label{lastpage}
\end{document}